\documentclass[12pt]{article}
\usepackage[tmargin=1in,bmargin=1in,lmargin=1.25in,rmargin=1.25in]{geometry}

\usepackage{setspace}
\usepackage{csquotes}

\usepackage{amsmath}
\usepackage{amsfonts}
\usepackage{stmaryrd}
\usepackage{dsfont}
\usepackage{graphicx}
\usepackage[title]{appendix}

\graphicspath{ {images/} }

\title{Smooth Infinitesimals in the Metaphysical Foundation of Spacetime Theories}
\author{Published in \textit{Journal of Philosophical Logic}}
\date{Lu Chen}
\begin{document}
\maketitle

 \noindent \textbf{Abstract.} I propose a theory of space with infinitesimal regions called \textit{smooth infinitesimal geometry} (SIG) based on certain algebraic objects (i.e., rings), which regiments a mode of reasoning heuristically used by geometricists and physicists (e.g., circle is composed of infinitely many straight lines). I argue that SIG has the following utilities. (1) It provides a simple metaphysics of vector fields and tangent space that are otherwise perplexing. A tangent space can be considered an infinitesimal region of space. (2) It generalizes a standard implementation of spacetime algebraicism (according to which physical fields exist fundamentally without an underlying manifold) called \textit{Einstein algebras}. (3) It solves the long-standing problem of interpreting   \textit{smooth infinitesimal analysis} (SIA) realistically, an alternative foundation of spacetime theories to real analysis  (Lawvere 1980). SIA is formulated in intuitionistic logic and is thought to have no classical reformulations (Hellman 2006). Against this, I argue that SIG is (part of) such a reformulation. But SIG has an unorthodox mereology, in which the principle of supplementation fails.

\vspace*{4mm}

\noindent\textbf{Keywords.} Continuum; smooth infinitesimal geometry; smooth infinitesimal analysis; vectorial quantity; tangent space; Einstein algebras; nonclassical mereology.

\section{Continua with Infinitesimal Parts}

What is a circle? Some---for example, Bryson of Heraclea, Kepler, Galileo and Leibniz---say that a circle is a regular polygon with infinitely many sides, each of which has an infinitesimal length (Boyer 1959; see also Bell 2017). This conception of a circle was utilized in reasoning about the area of a circle: just as the area of a regular polygon with finitely many sides is equal to half of the product of its apothem (the distance from the center to a side) and its perimeter, the area of a circle as a regular polygon with infinitely many sides is equal to half of the product of its radius and its perimeter. In general, any smooth curve is composed of infinitesimally short straight microsegments.

This idea about circles and curves also continues to be part of widely used heuristic reasoning in physics. For example, consider an object rotating along a circular orbit at a constant speed. What is its acceleration? Here's how physicists use infinitesimals to anticipate the answer. We can first imagine that the object in question rotates for an infinitesimal amount of time $\Delta t$. By assuming (among other things) that the trajectory of the object during $\Delta t$ is straight, we can easily obtain the correct equation for its acceleration (see Morin 2008).

But this mode of reasoning is considered merely heuristic, because space and time (or spacetime) standardly construed do not have infinitesimal parts. \textit{The standard view} says that a line is composed of uncountably many unextended points, which can be algebraically represented by real numbers. Every extended part of a line has a finite length, and there are no infinitesimal line segments. The aforementioned reasoning can be regimented by the limit approach without involving infinitesimals. For example, velocity is the limit of distance divided by time as the distance in question gets smaller and smaller. As this view became more influential, infinitesimals were condemned as “cholera-bacilli” (Cantor 1887) and “unnecessary, erroneous, and self-contradictory” (Russell 1903).

However, the standard view faces many conceptual difficulties. For one thing, the reformulation of infinitesimal reasoning in terms of limits is rather baroque and unintuitive, and leads to various interpretative issues for vectorial-like physical quantities.  If a velocity is the limit of distance divided by time as the time approaches zero, then is it a mere logical construction out of occupying various spacetime points (for example, see Tooley 1988, Arntzenius 2000, Butterfield 2006)? The nature of the electromagnetic field is similarly puzzling: are the vectorial field values extrinsic to spacetime points (see Weatherson 2006, Busse 2009)? Things get even messier when it concerns curved spacetime, which involves tangent spaces and their relations. Are tangent spaces physical spaces over and beyond our ordinary spacetime, or are they reducible to other physical features of the world? How are different tangent spaces joined up in curved spacetime? Reading answers directly from the mathematical formalism would commit us to all kinds of abstract entities at the fundamental level.

These questions (among others) motivate us to look for alternatives to the standard view. In this paper, I would like to address these questions by advancing a novel theory of space with infinitesimal regions and consider tangent space as such regions.   Although infinitesimals were banished from standard calculus, researches on infinitesimals have been vibrant (see Ehrlich 2006). I will utilize the mathematical literature on an alternative foundation of calculus called \textit{smooth infinitesimal analysis} (SIA) (part of the more encompassing theory \textit{synthetic differential geometry} (SDG)), developed by Lawvere and others (see Lawvere 1980, Kock 1981, Moerdijk and Reyes 1991).  (Another well-known infinitesimal theory is Robinson's (1966) \textit{nonstandard analysis}, which augments the familiar real numbers with infinite numbers and their infinitesimal inverses. But this theory is less relevant to the geometric considerations I have mentioned.\footnote{Here's a brief explanation for why. According to nonstandard analysis, we can use the same arithmetic operations (e.g., multiplication, square root, logarithm) on the infinitesimals as on standard real numbers. A polygon with $N$ sides---where N is an infinite number in nonstandard analysis---is not a circle but still a polygon with sharp angles ($180^\circ-\frac{360^\circ}{N}$). (see Bell 2008, Mayberry 2000, Reeder 2015; for infinitesimal theories of space based on nonstandard analysis, see Chen 2019, 2020)})

While SIA is an interesting candidate because it is motivated exactly by the purpose of regimenting physicists' mode of reasoning with infinitesimals (see Lawvere 1980, Moerdijk and Reyes 1991), there is a serious problem with interpreting its statements literally and realistically. The problem is that it is formulated in intuitionistic logic and is classically inconsistent (see Bell 2008). Hellman (2006) also argued that there SIA cannot be reformulated as a classically consistent theory of infinitesimals.\footnote{There is a disagreement on the significance of this problem in the literature. As I will also mention later in the paper, some people hold that we should give up on classical logic as the correct logic for fully-interpreted theories (Heller and Kr\'ol 2016), while others hold that we should give up SIA as a candidate realistic theory of space (Hellman 2006, Reeder 2015).} 

But there is a way for classical logicians to share the insight of SIA, for I will argue that there is a classically consistent theory of space with infinitesimal regions in its vicinity that also regiments the aforementioned mode of reasoning. I will advance such a theory---\textit{smooth infinitesimal geometry} (SIG)---based on the models for SIA proposed by Moerdijk and Reyes (1991) consisting of algebraic objects called \textit{rings} (Section 2). According to SIG, each of those rings represents a region of space (or spacetime) and some represent infinitesimal ones. I will argue that this new theory has several distinct merits. First of all, it can provide a straightforward understanding of vectorial physical quantities and tangent spaces (Section 3). We can indeed say that, for example, velocities are properties of trajectories that are intrinsic to infinitesimal durations of time. Similarly, electromagnetic field values are intrinsic properties of infinitesimal parts of the field. A tangent space at a spacetime point is simply an infinitesimal spacetime region. However, I will leave open whether this theory is overall better than the standard view due to its very unusual features: its mereology is nonclassical in that the principle of supplementation (according to which X is a proper part of Y only if Y has a proper part disjoint from X) fails. My underlying interest is to expand our understanding of space with conceptually rigorous, mathematically grounded, and technically fruitful alternative theories of space.

Second, SIG  generalizes \textit{Einstein algebras} discussed in philosophy of physics, which is considered a distinct way of conceptualizing relativistic physics alternative to substantivalism (see Geroch 1972, Earman and Norton 1987, Rosenstock, Barrett and Weatherall 2015, Menon 2019). The algebraic objects involved in this approach typically constitute a proper subcollection of those involved in SIG because infinitesimal regions are excluded.\footnote{The \textit{explicit} exclusion, as far as I am concerned, only occured in Rosenstock et al (2015). It may be excluded in other authors' work as a non-obvious consequence of their formalism. But this requirement is not necessary for doing physics, as demonstrated in Chen and Fritz (2021). } Such an exclusion, as I will argue, is unnecessary. 

I will further argue that SIG can be considered a realistic interpretation of SIA and thereby overcome the main interpretative difficulty of SIA (Section 5).  Recall that SIG is based on the classical models for SIA. The usual perspective on the relation between SIA and its models is that the models are abstract structures invoked to prove the intuitionistic consistency of SIA (Bell 2008, Hellman 2006). But we can reverse the usual order of things:  instead of taking the object language realistically and its models instrumentally, we can take the models realistically and the object language instrumentally (akin to the semantic view of theories in van Fraassen 1980). Also, an interpretation ought to preserve the important virtues of the theory to be interpreted. SIG indeed satisfies this and therefore captures the realistic significance of SIA.

It might be worth mentioning that there are other recent proposals for alternative theories of nilpotent infinitesimals that obey classical logic, such as Giordano's (2010) ring of Fermat reals. These alternative approaches---while deserving more attention---will not be discussed in this paper. But I shall briefly note that these alternative theories do not have the distinct features of SIG. For example, Giordano's approach does not provide a unique interpretation of a tangent space.\footnote{In Giordano's approach, a line is not represented by real numbers but by $a+bx$, where $a,b$ range over real numbers, and $b^2=0$. Then, for each real number $a$ (standard point), it is associated with infinitely many infinitesimal segments $[a-bx,a+bx]$, each of which seem to have equal standing to represent a tangent space. In contrast, the approach I am considering features a unique ``smallest'' extended region around each point that serves as a tangent space (``smallest'' in the sense that it is properly contained in all other extended regions around the point). } It also does not have a natural connection to Einstein algebras.

\section{Smooth Infinitesimal Geometry}

In this section, I will present a preliminary theory of space with infinitesimal regions that is useful for regimenting geometricists' and physicists' mode of reasoning with infinitesimals and for solving the conceptual difficulties related to vectorial quantities and tangent space. I call this theory \textit{Smooth Infinitesimal Geometry} (SIG).

\subsection{Ontology and Mereology}

 According to SIG, all regions of space can be represented by \textit{rings} of smooth functions on real coordinate space (``smooth'' means being indefinitely differentiable).  A ring is a set closed under binary operations \textit{addition} and \textit{multiplication} satisfying certain axioms like distributivity.  Smooth functions are closed under addition and multiplication. For example, if we multiply $f(x)=ax$ with $g(x)=b$, we get $h(x)=abx$.  For simplicity, I will start by focusing on one-dimensional region of space (or pretending that our space is one-dimensional).  I will postulate that regions of (one-dimensional) space are represented by all the \textit{quotient rings} of $C^\infty(\mathbb{R})$, the ring of all smooth functions  over the real line $\mathbb{R}$.

A \textit{quotient ring} of $C^\infty(\mathbb{R})$ consists of equivalence classes of members of $C^\infty(\mathbb{R})$ under certain equivalence relations that preserve the original ring structure. For example, consider the equivalence relation of having the same value on the real interval $[0,1]$. Then, two smooth functions belong to different equivalence classes if and only if their values differ on $[0,1]$, which allows each class to be represented by a smooth function on $[0,1]$.  Thus, these equivalence classes form a quotient ring isomorphic to the ring of all smooth functions on $[0,1]$, which is denoted by $C^\infty([0,1])$.\footnote{In general, for any closed set $A$ of real numbers, the set of all equivalence classes consisting of all smooth functions that agree on $A$ constitutes a quotient ring of $C^\infty(\mathbb{R})$ isomorphic to the ring of all smooth functions on $A$. Note, however, that the set of all smooth functions on an open set of real numbers does not constitute a quotient ring of $C^\infty(\mathbb{R})$. Consider $C^\infty((0,1))$. This includes smooth functions whose extensions to [0,1] diverge at point 0 and 1. Such functions cannot be extended to smooth functions on $\mathbb{R}$ and thus do not correspond to any equivalence class of smooth functions on $\mathbb{R}$ as previously indicated. However, this is a quotient ring of $C^\infty(\mathbb{R}^2)$. In fact, for every manifold, the set of all smooth functions on it is a quotient ring of $C^\infty(\mathbb{R}^n)$ for some dimension $n$.} (Henceforth I will identify quotient rings with such representatives for brevity.) Similarly, $C^\infty(\{0\})$, or the ring of real numbers $\mathbb{R}$, is a quotient ring of $C^\infty(\mathbb{R})$ under the equivalence relation of having the same value at zero.

Furthermore, I postulate that the quotient relation between rings represents the parthood relation between regions. For example, since $C^\infty([0,1])$ is a quotient ring of $C^\infty(\mathbb{R})$, the region it represents is a part of the region represented by $C^\infty(\mathbb{R})$. It is not hard to see, then, that $C^\infty(\mathbb{R})$ represents the whole space, while its proper quotient rings represent its proper parts. To take stock, we have the following principle:

\begin{quote}
	\textsc{Ring Representation}. There is a one-to-one correspondence between all nonzero quotient rings of $C^\infty(\mathbb{R})$ and all regions of space such that region X is a part of region Y iff X's corresponding ring is a quotient ring of Y.\footnote{I admit only nonzero rings because the zero ring would correspond to the null region, which I do not consider real.}
\end{quote}
Before I expound on the implications of this principle, let me first contrast it to \textit{the standard view} of space. According to the standard view, (one-dimensional) space consists of points that can be algebraically represented by real numbers. Every region can be represented by a subset of the real line, and one region is a part of another if the corresponding set of the former is a subset of that of the latter. Some readers might wonder whether \textsc{Ring Representation} eventually amounts to an equivalent view of space with merely different representations: instead of representing a region with a subset of the real line, it might seem that I represent it with all the smooth functions on the subset (for any two closed subsets $A,B$ of $\mathbb{R}$, if $A$ is a subset of $B$, then $C^\infty(A)$ is a quotient ring of $C^\infty(B)$). What is the difference? Let me explain.

To reason about the relations between regions more conveniently in the upcoming discussions, I shall utilize the technical notion of \textit{ideals}  which uniquely determine quotient rings. An \textit{ideal} of a ring is a subset of it closed under addition and multiplication by ring members. For example, the set of even numbers is an ideal of the ring of natural numbers because $even + even = even$ and $even \times integer = even$.  It is helpful to think of an ideal of a ring as a set of elements that can be consistently identified with zero---it is ``collapsable'' set. More technically, for any rings $A$ and $B$, and any homomorphism from $A$ to $B$, the set of all elements that it maps to zero in $B$ is an ideal of $A$ (a \textit{homomorphism} is a map that preserves the ring structure).  For example, the set of all smooth functions that vanish at zero is an ideal of $C^\infty(\mathbb{R})$  (call it ``$I_0$") because those functions can be mapped to zero in $\mathbb{R}$ under the map that takes each function to its value at zero. Every ideal of a ring defines a quotient ring by collapsing all elements that agree within that set, just as $I_0$  defines the ring $\mathbb{R}$ by collapsing all functions that agree at zero.

A quotient ring that plays a special role in SIG is the ring of \textit{affine functions} on $\mathbb{R}$, which have the form $f(x)=a+bx$. Call this ring $\mathfrak{L}$ (for ``linear''). This is a quotient ring of $C^\infty(\mathbb{R})$ under the equivalence relation of having the same value at zero \textit{and} the same derivative at zero. In this case, each equivalence class can be represented by an affine function with its coefficients respectively being the value and the derivative of the members of that equivalence class at zero.\footnote{As a result, addition on $\mathfrak{L}$ is as usual, but multiplication on $\mathfrak{L}$ is the ``affine approximation'' of the multiplication on $C^\infty(\mathbb{R})$---that is, for two affine functions $f$ and $g$, if $h$ is their product in $C^\infty(\mathbb{R})$, then their product in $\mathfrak{L}$ equals $h(0)+h'(0)x$). A special feature of the ring is that it contains \textit{nilpotent elements}, elements that square to zero. For example, consider the function $f(x)=bx$, where $b$ is an arbitrary real number. Since $f^2(x)=b^2x^2$ has zero value and zero derivative at zero, it equals the zero function in $\mathfrak{L}$. So $f(x)=bx$ is a nilpotent element of $\mathfrak{L}$.} The ideal that determines this ring is the set of all smooth functions that have both zero value and zero derivative at zero  (call it ``$I_\Delta$").

Because ideals uniquely determine quotient rings, \textsc{Ring Represention} implies that every region of space can be represented by an ideal of $C^\infty(\mathbb{R})$. The parthood relation between regions can be conveniently represented by the subset relation between their corresponding ideals, but with the direction \textit{reversed}. This is because the larger the ideal is,  the fewer equivalence classes it yields, which form a smaller quotient ring, and therefore the smaller region it represents. For example, $\mathbb{R}$ is  a quotient ring of $\mathfrak{L}$, and correspondingly, we have $I_0\supset I_\Delta$. To highlight, \textsc{Ring Representation} implies: 

\begin{quote}
	\textsc{Ideal Representation}. A region $X$ is a part of another region $Y$ if and only if $Y$'s representing ideal is a subset of $X$'s ideal. 
\end{quote}

Now, I can show a few basic results about the structure of space based on this principle. First:

\begin{center}\smallskip
	\emph{Space contains points that are mereologically simple.}\smallskip
\end{center}
To show this, we just need to show that $C^\infty(\mathbb{R})$ have maximal (nontrivial) ideals. Indeed, $I_0$ is one such ideal. Suppose there is a larger ideal $I$ than $I_0$. Let $f$ be an element of the larger ideal that is not in $I_0$. Then, $f$ is a smooth function whose value at zero is not zero. Then, every smooth function over $\mathbb{R}$ can be obtained through $f$, which means that this ideal equals the whole ring $C^\infty(\mathbb{R})$, and therefore trivial.\footnote{Suppose $f(0)\neq 0$. Then for any smooth function $g$, we have $$g=\frac{g(0)}{f(0)}f+(g-\frac{g(0)}{f(0)}f)$$ Here, $\frac{g(0)}{f(0)}f$ is in $I$ because $f$ is in the ideal and $\frac{g(0)}{f(0)}$ is a ring element. $g-\frac{g(0)}{f(0)}f$ is also in $I$ because it vanishes at zero and therefore is in $I_0\subset I$.  So, $g$ is in $I$.} Therefore, $I_0$ is a maximal (nontrivial) ideal. We say, then, for any real number $p$, the ideal of all smooth functions that vanish at $p$ represents a point $p$ in space. 

Second:

\begin{quote}\smallskip
	\emph{Each point is contained in an infinitesimal region that is part of all other regions around the point.}\smallskip
\end{quote}
Putting aside ``infinitesimal'' for the moment, this claim amounts to there being next largest ideals to the maximal ideals in the sense that there no other ideals strictly between them and the maximal ideals. We can show that $I_\Delta$ is indeed the next largest ideal to $I_0$.\footnote{One may ask whether there are ideals that neither properly include $I_\Delta$ nor are properly included by $I_\Delta$. Yes, there are. But such ideals do not represent regions \textit{around} point zero. For example, the ideal of all functions that vanish at both zero and some other point represents the fusion of zero and the other point. This ideal is not strictly between $I_0$ and $I_\Delta$.}  Suppose there is an ideal $I$ between them. Let $f$ be an element of $I$ that is not in $I_\Delta$. Then, $f$ is a smooth function that either has a non-zero value or a non-zero derivative at zero. By the previous reasoning, $f$ cannot have a non-zero value at zero without $I$ becoming trivial. But by a similar reasoning, $f$ cannot have a non-zero derivative either without $I$ being identical to $I_0$.\footnote{Suppose $f'(0)\neq 0$. Then for any smooth function $g$ that vanishes at zero, we have $$g=\frac{g'(0)}{f'(0)}f+(g-\frac{g'(0)}{f'(0)}f)$$ This shows that $g$ is also in $I$ because both $\frac{g'(0)}{f'(0)}f$ and $g-\frac{g'(0)}{f'(0)}f$ are in $I$ (the latter is in $I_\Delta$). Thus, $I=I_0$. } Thus, the supposition is impossible. So, $I_\Delta$ is the next largest ideal, and represents the smallest region around zero. I will call it \textit{a nilpotent region}. Note that the region is so minuscule that it does not contain any other point. We can picture each point of space as tightly ``shelled'' by such nilpotent regions. Moreover, we can show that such a shell is further shelled by a third smallest region, which in turn is shelled by a fourth smallest, \textit{ad infinitum}. Each of these infinitely many layers can be represented by the ideal of smooth functions whose zeroth, first, ..., and $n$th derivatives all vanish at some point for some appropriate $n$. 

All these regions are infinitesimal in the sense that they are contained in every finite interval around the relevant points. A finite interval region $[a,b]$ is represented by the ideal of all smooth functions that vanish on the interval $[a,b]$. It is easy to confirm that all infinitesimal regions around a point are contained in all the finite intervals around that point: if a function vanishes at an interval, then its $i$-th derivative at a point properly within the interval has to be zero for any $i$.  Another way of seeing these regions as infinitesimal is that their ideals throw out all information of a smooth function other than its value and derivatives at a point.

To summarize: space is composed of unextended points, each of which is contained in an infinitesimal region (and indeed infinitely many of them) that is properly contained in all finite regions around it. A nilpotent region is the smallest infinitesimal region around a point, which has the point as its only proper part. 

\paragraph{Nonclassical mereology}

The above picture implies that the mereology of space does not satisfy classical mereology. In particular, the principle of \textit{supplementation}, according to which there is always a reminder when you ``subtract'' a proper part from a whole. 

First, we can quickly confirm that the mereology of space satisfies the following three core axioms of parthood (Varzi 2019):

\begin{quote}
	\textsc{Reflexivity}. Every region is a part of itself.
	
	\textsc{Transitivity}. If region $X$ is a part of region $Y$, and $Y$ is a part of region $Z$, then $X$ is a part of $Z$.
	
	\textsc{Antisymmetry}. No two distinct regions are part of each other.
\end{quote}
\noindent These three axioms are deemed essential to any mereology, so it is a reassurance that the mereology postulated indeed satisfies them. They are satisfied because the subset relation between ideals is reflexive, transitive and antisymmetric.

However, the mereology of space violates supplementation, which is an important (but not core) axiom of classical mereology:

\begin{quote}
	\textsc{Supplementation.} If $x$ is a proper part of $y$, then $y$ also has a proper part $z$ that is disjoint from $x$.
\end{quote}
This is violated because a point is a proper part of a nilpotent region, which has no other proper parts disjoint from the point. Moreover, a nilpotent region is contained in a third smallest region which does not have other disjoint parts. So on and so forth. We can see that this picture differs drastically from the standard view as well as alternative infinitesimal theories of space. For example, in a theory of space based on nonstandard analysis proposed by Chen (2020), there are no smallest infinitesimal regions---every infinitesimal interval contains a smaller interval, and there is no violation of supplementation.

To be clear, I do not intend to defend this violation, but will leave it to the readers to weigh this cost against the benefits of SIG, which I will explain in the upcoming sections.  Instead, I shall \textit{contrast} this feature with other violations of supplementation in the literature that I am aware of (see Varzi 2019).  For example, some philosophers argue that a lump of clay that constitutes a statue is a part of the statue but not identical to it, since they have different modal properties (the lump of clay can survive squashing while the statue can't). However, there are no other parts that make up the difference between them (Thomson 1998, Walters 2017). So, supplementation is violated. But the counterexample in our case has a very different feature. The statue and its lump of clay are spatially coextensive and geometrically indiscernible, while a nilpotent region and a point are not spatially or geometrically identical. The idea that the mereology of \textit{space} could violate \textsc{Supplementation} has rarely been discussed (with the exception of Forrest 2003, 2007).\footnote{The violation of supplementation also leads to other strange features. For example, suppose we define ``fusion" in the usual way, namely that $x$ is a fusion of $y$s just in case $y$s are parts of $x$ and everything that overlaps with $x$ also overlaps with one of $y$s.  Also suppose that the fusion of all nilpotent regions, each of which includes a  point, is the whole line. Then, it follows that the fusion of all points is also the whole line, even though intuitively they do not cover any nilpotent region.  Forrest (2007) discusses alternative definitions of ``fusion'' when \textsc{Supplementation} is violated.} 

While the loss of supplementation is counterintuitive, I believe that we should not reject the theory outright on this ground. For one thing, our intuition about the structure of reality can be highly unreliable when it comes to scales wildly out of the ordinary ranges, especially when it comes to infinities and infinitesimals.  It is well accepted that infinities are bizarre---for example, the Hilbert hotel is full but can accommodate infinitely more guests. Even for finitely small scales, our intuitions become flimsy, as witnessed by highly counterintuitive quantum physics. 

\subsection{Differential Structure and Vectors}

I will introduce the differential structure of space as well as the notion of vectors, in preparation for the metaphysical explication of physical vectorial quantities such as velocity and electromagnetic field values (in general, structures involving tangent spaces) in the next section. I will first focus on one-dimensional space, and extend to the higher-dimensional case toward the end.

To postulate the differential structure of space, we need to first postulate smooth functions on space. Those functions should be considered primitive structures that are not definable through point sets---after all, a nilpotent region contains no more than a single point but should admit more functions than a single point does. We shall let homomorphisms between quotient rings of $C^\infty(\mathbb{R})$ represent smooth maps between regions of space. In particular, for any region represented by a quotient ring, let homomorphisms from $C^\infty(\mathbb{R})$ to that quotient ring represent \textit{smooth functions} on this region.

More precisely, let $C^\infty(\mathbb{R})/I$ be the quotient ring of $C^\infty(\mathbb{R})$ determined by ideal $I$, which consists of equivalence classes whose members agree within the ideal. Let's strengthen \textsc{Ideal Representation} with the following clause :

\begin{quote}
	\textsc{Smooth.} For any ideal $I$ of $C^\infty(\mathbb{R})$, all smooth functions on the region represented by $I$ can be represented by members of $C^\infty(\mathbb{R})/I$. (That is, the ring of smooth functions on the region is isomorphic to $C^\infty(\mathbb{R})/I$.)
\end{quote}
For convenience, I will identify an equivalence class of smooth functions with ``the common part'' of those functions whenever possible. For example, the equivalence class of all smooth functions that have the same value on zero is identified with their value on zero. Similarly, the equivalence class of all smooth functions that have the same values on the real interval $[0,1]$ is identified with their restriction to $[0,1]$ (which is a smooth function only defined on $[0,1]$). It follows from \textsc{Smooth} that all the smooth functions on space can be represented by smooth functions on the real line (in this case, the ideal in question is the zero ideal consisting of only the zero function). Also, all the smooth functions on a nilpotent region can be represented by   affine functions. In this sense, all smooth functions on a nilpotent region are affine.\footnote{This is very similar to \textsc{Kock-Lawvere Axiom}, a central axioms in SIA (smooth infinitesimal analysis). See Section 5.}

	 A \textit{trajectory} is a path that an object follows through space as a function of time. In standard formalism,  it is usually expressed as a function from the unit real interval $[0,1]$ to real-coordinate space. Suppose time has the same structure as one-dimensional space. Then, \textsc{Smooth} entails that a trajectory in one-dimensional space can be represented by a smooth function on the real interval $[0,1]$, just like in standard formalism.

The differential structure of space turns out to be attractively simple. The \textit{derivative} of a smooth function $f$ at a point $x$ is simply the slope of the affine function resulting from \textit{restricting} $f$ to the nilpotent region around $x$. But first, let me explain what ``restricting a smooth function'' means in our picture given \textsc{Ideal Representation} and \textsc{Smooth}. Let the representation function that maps ideals to regions be denoted ``$R$'': that is, let the region represented by an ideal $I$ be $R(I)$. Let ``$[f]_I$'' refer to the equivalence class defined by ideal $I$ with representative $f$.

\begin{quote}
	\textsc{Restriction.} For any ideals $I$ and $J$, if $J\subseteq I$, then for any function on region $R(J)$ represented by $[f]_J\in C^\infty(\mathbb{R})/J$, the \textit{restriction} of the function to region $R(I)$ is represented by $[f]_I\in C^\infty(\mathbb{R})/I$. 
\end{quote}

\noindent This definition is not arbitrary. Restricting a function on $R(J)$ to a smaller region $R(I)$ amounts to abandoning information outside region $R(I)$, which in turn amounts to ignoring the difference between functions according to the broader equivalence relation determined by $I$ that defines the smaller region.  For example, consider restricting a function on space represented by a smooth function $f$ on $\mathbb{R}$ to region $[0,1]$. Recall that the region $[0,1]$ is represented by the ideal of all functions vanishing at the real interval $[0,1]$. It follows from \textsc{Restriction} that the restriction is represented by the equivalence class $[f]$ consisting of all smooth functions agree with $f$ on $[0,1]$, which we conveniently identify with $f$ restricted to the real interval $[0,1]$.  This means that the restriction of the function represented by $f$ to region $[0,1]$ just amounts to the restriction of $f$ to the real interval $[0,1]$ in the standard sense.

It gets more interesting when it comes to restricting a function to a nilpotent region. Consider the nilpotent region $R(I_\Delta)$. The restriction of a smooth function to this region amounts to ignoring all information other than its value and its derivative at zero. More elaborately, for any smooth function on space represented by smooth function $f$ over $\mathbb{R}$, its restriction to the nilpotent region is represented by the affine function $g(x)=f(0)+f'(0)x$. This applies to other nilpotent regions as well: for any point $p$, the restriction of the function represented by $f$ to the nilpotent region around $p$ is represented by $g(x)=f(p)+f'(p)x$. This implements the idea that, for any smooth curve and any point on the curve, there is a straight microsegment of the curve around that point. This is how we can make sense of the mode of reasoning with infinitesimals employed by geometricists and physicists (Section 1).  For any affine function represented by $f(x)=a+bx$, call $b$ the \textit{slope} of the function. Then, it is natural to define the \textit{derivative} of a smooth function at a point $p$ to be the slope of the function restricted to the nilpotent region around $p$.

\begin{quote}
	\textsc{Derivative.} For any smooth function $f$ on a region $X$, and for any point $p$ in $X$, the \textit{derivative} of $f$ is the slope of the restriction of $f$ to the nilpotent region around $p$.
\end{quote}

Similarly,  second-order derivatives of a smooth function at various spacetime points are intrinsic to second smallest infinitesimal shells around those points (namely regions represented by ideals of all functions that have zero value, zero first derivative, and zero second derivative at those points). Second-order derivatives are ubiquitous in physics. The Laplace operator, for example, is a second-order derivative operator in the one-dimensional case.

The strategy presented so far also applies to the higher-dimensional cases. A detailed account of such cases is beyond the scope of the paper, but I will give a flavor of it here. Consider the $n$-dimensional case. We can let regions be represented by ideals of $C^\infty(\mathbb{R}^n)$. In particular, a nilpotent region can be represented by an ideal of smooth functions over $\mathbb{R}^n$ that have zero value and zero derivative at a point. The resulting quotient ring is isomorphic to the ring of affine functions over $\mathbb{R}^n$. Such affine functions can be more generally expressed as $f(\textbf{x})=a+\textbf{bx}$, where \textbf{x} ranges over $n$-dimenisional vectors (or ``$n$-vectors''), and \textbf{b} is a \textit{covector} (also known as ``dual vector'' or ``1-form'') which maps $n$-vectors to real numbers. The differentiation works very similarly as the one-dimensional case. The restriction of a smooth function $f$ on an n-dimensional region to a nilpotent region around point $p$ corresponds to two values: one is a real number, which is the value of $f$ at $p$, and the other is a covector, which defines the \textit{total derivative} or the \textit{gradient} of $f$ at $p$.\footnote{Covectors (which are associated with derivatives) and vectors (which are associated with gradients) are closely related. In Euclidean space, covectors and vectors are equivalent. In other cases, covectors can be easily converted to vectors given a pseudo-Riemannian metric. Metrics are not discussed in this paper, which can raise complicated questions for SIG and can be discussed elsewhere. For this paper, I shall pretend that space is Euclidean.} As I will explain in the next section, this provides a straightforward conceptual foundation for vectorial physical fields.

\section{Metaphysics of Vectors}

One of the main advantages of SIG is that it offers a straightforward metaphysical analysis of vectorial-like physical quantities or objects that otherwise lack an attractive account. Take a simple vectorial quantity, velocity, as an example. Standardly, the velocity of an object following trajectory $q$ at time $t$ is the derivative of $q$ at $t$, which in turn is understood as a certain limit as we approach smaller and smaller parts of the trajectory around $t$. There has been a lasting debate on whether velocity is intrinsic to an instant of time (Butterfield 2006). It seems wrong to consider velocity intrinsic to an instant because the velocity of an object does not only depend on its position at the instant but also on its positions at nearby times (the nearby trajectory needs to approach a certain limit). But the standard view, which reads the limit definition of velocity literally and reduces velocity to occupying various spacetime points, also seems unattractive. For one thing, it causes difficulties for formulating determinism and precludes velocity from being part of an explanation for motion (Tooley 1988, Arntzenius 2000). 

But in our picture, where space and time are enriched with infinitesimal parts, we have a simple solution. Under the definition of derivatives in SIG, the velocity in question is simply the slope of the trajectory $q$ restricted to the infinitesimal duration $\Delta_t$ around $t$.  The trajectory of the object during $\Delta_t$ is characterized by a pair of real numbers, one of which indicates the location of the object, and the other indicates its velocity. Note that the velocity of the object during the infinitesimal duration does not depend on or impose any restriction on any part of the trajectory outside that duration. Hypothetically, even if an object existed only for an infinitesimal duration, it still would have a well-defined velocity.  Therefore, we can say that velocity is intrinsic to an infinitesimal duration. This avoids the difficulties that other views face. It would be natural, for example, to formulate determinism in terms of the intrinsic states at instants taken as infinitesimal durations.

Second, consider the electric field in three-dimensional space, which is a vector field. The physical meaning of the value of an electric field at a point is the force (per unit charge) a charged body would experience at that point. Thus, the electric field value at a point has a spatial direction that determines the trajectory of a charged body passing through that point. Like the case of velocity, there is a debate on whether we should consider those vectors as intrinsic to spatial points (see Weatherson 2006, Busse 2009). If those vectors are intrinsic to the electric field within spatial points, then it is strange that they can have spatial directions since a point does not have any spatial direction. But if they are extrinsic to points, then it is unclear how we should think about them. While velocity is standardly reduced to occupying certain positions at various times, it is unclear what electric field values are reduced to or whether they should be reduced at all (we can also consider an electromagnetic field instead, which is considered fundamental). Once again, SIG provides an easy solution.    We can say that the electric field values are properties of the electric field  intrinsic to infinitesimal regions, or simply that they are infinitesimal parts of the field.\footnote{Or we can say that the electric field values are properties of the \textit{electric potential field}, which is a scalar field of which the electric field is the gradient. Recall that when restricting a scalar field $f$ to a nilpotent region around $p$, we obtain the value and the gradient of $f$ at $p$ (Section 2.2). So restricting the electric potential field to a nilpotent region around a point results in the electric field value at that point.} 

More generally, we now have a new metaphysical framework for understanding physical structures based on tangent spaces and tangent bundles (see Arntzenius and Dorr 2012). Informally, a tangent space at a point encodes all directions one can pass through that point. The nature of such spaces is puzzling: it's unclear whether they are physical objects (perhaps spaces) in addition to our ordinary space or reducible to other physical features of the world. Like vectorial quantities, it is puzzling whether they are intrinsic features of points. Moreover, how are they joined up in various ways as in curved spacetime? How should we understand curvatures and connections? Directly reading the answer from the mathematical formalism would require us to posit all sorts of platonic objects which have mysterious interactions with the physical world.  But a theory of space with nilpotent regions starts to suggest answers to these questions based on a simple explication of tangent space. Since all the smooth functions of the form $f(\textbf{x})=\textbf{bx}$ on a nilpotent region form a vector space, a nilpotent region is just like a tangent space. In this way, we can regiment the idea that tangent spaces are infinitesimal parts of physical space, and pave the way for understanding more complicated infinitesimal structures.

\section{Generalizing Einstein Algebras}

The idea of using algebraic structures like rings to represent geometric entities is not new. Indeed, algebraic geometry is a prominent branch of mathematics that studies the relation between geometric and algebraic structures. In this section, I will explain how SIG generalizes and expands an established way of conceptualizing space in the literature of philosophy of physics.

\textit{Spacetime algebraicism}, the view that physical fields exist fundamentally without an underlying spacetime, appeals to algebraic structures of smooth functions as substitutes for space. This view is initially proposed as a way to avoid the difficulties troubling spacetime substantivalism (see Geroch 1972, Earman and Norton 1987). However, Rynasiewicz (1992) forcefully points out that there are parallel difficulties against such an approach.\footnote{Rynasiewicz's conclusion is contended by Bain (2003), who argues that the difficulties are not exactly parallel.} Although this approach does not achieve some of the goals it was set out for, it provides an alternative framework of formalizing space and geometry with various possible implementations that have other advantages (see Bain 2003, Chen and Fritz 2021, Chen 2022).\footnote{Many alternative formalisms of spacetime algebraicism are underexplored, which may have advantages over the substantivalist approach to spacetime. For example, non-commutative geometry (which is associated with noncommutative algebras of functions) is an important approach to physics, but it is unclear what structure of space underlies those noncommutative functions (see Bain 2003 and Heller and Sasin 1999). } In what to follow, I will explain the connection between SIG and a standard implementation of spacetime algebraicism called \textit{Einstein algebras}, and argue that SIG generalizes the latter. 

Our physical space is standardly modeled by smooth manifolds in standard differential geometry. Roughly, \textit{manifolds} are topological spaces that are locally like a real coordinate space $\mathbb{R}^n$, and a \textit{smooth} manifold have a differential structure defined by smooth functions. Here's the basic idea of spacetime algebraicism: every smooth manifold $M$ can be substituted by an algebraic structure $M*$ consisting of all the smooth functions on the manifold---called \textit{Einstein algebra}---as all information of the manifold is preserved.\footnote{More precisely, an Einstein algebra has an additional metric structure that corresponds to the metric of a Lorentzian manifold, which is used to model space in general relativity (see Rosenstock et al 2015: 310, 314). But I ignore the metric aspect in this paper.}  For example, given two smooth manifolds $M,N$, a smooth map $\phi$ from $M$ to $N$ can be uniquely encoded by a homomorphism from $N*$ to $M*$ that maps each smooth function on $N$ to its composition with $\phi$, which is a smooth function on $M$. The other direction also holds: all information about Einstein algebras can be encoded in smooth manifolds.

\begin{quote}
	\textsc{Manifold-Algebra Duality.}
	
	There is a one-to-one correspondence between smooth manifolds and Einstein algebras such that for any two manifolds $M$ and $N$, every smooth map from $M$ to $N$ uniquely corresponds to a homomorphism from $N*$ to $M*$ and vice versa. (Rosenstock et al.\@ 2015)
\end{quote}
This one-to-one correspondence between manifolds and smooth algebras is called a ``duality''  because maps between manifolds correspond to homomorphisms between algebras with the \textit{opposite} directions.

Given a manifold $M$, an Einstein algebra is a special sort of quotient ring of $C^\infty(M)$. Not all quotient rings of $C^\infty(M)$ are isomorphic to the rings (or algebras) of all smooth functions on some manifolds.\footnote{For our purposes, the difference between rings and algebras is not important.} In particular, not all quotient rings of $C^\infty(\mathbb{R})$ are isomorphic to the rings of all smooth functions on some subspaces of $\mathbb{R}$.\footnote{As will become apparent soon, this is not entirely due to that $n$-manifolds are standardly defined as being locally like $\mathbb{R}^n$ and therefore a closed interval is not a manifold (since its boundary points do not have line-like neighborhoods), which entails that rings like $C^\infty([0,1])$ are not smooth algebras. The discussion in the main text won't be affected even if we use ``manifold" in the broad sense that includes those with boundaries.} For example, there are no (non-zero-dimensional) subspaces on which all smooth functions are affine. So the ring of affine functions $\mathfrak{L}$ does not correspond to any manifold. Indeed, Einstein algebras in Rosenstock et al.\@ (2015) are defined in a way that precisely excludes those rings like $\mathfrak{L}$ that do not correspond to any manifolds to ensure \textsc{Manifold-Algebra Duality}. More specifically, all rings that have non-zero nilpotent elements are ruled out for not being ``geometric," a defining condition for Einstein algebras (Rosenstock et al.\@  2015, 311).\footnote{An algebra $A$ is \textit{geometric} iff there are no non-zero elements of $A$ that belong to the kernels of all homomorphisms from $A$ to $\mathbb{R}$ (where a kernel of a homomorphism is the set of elements it maps to zero). For any homomorphism from a ring $A$ to $\mathbb{R}$, since it preserves squares of elements, all nilpotent elements in $A$ are mapped to 0. So if $A$ has non-zero nilpotent elements, then $A$ is not geometric.}

Note, however, that the condition of ``geometricity'' imposed on quotient rings has no independent justification apart from that it is a necessary condition for the rings to have the one-to-one correspondence with manifolds under standard differential geometry. But we are precisely considering alternatives to standard differential geometry, which should not be ruled out from the outset. (Insofar as spacetime algebraicism aims at dealing away with manifolds and substantive spacetime, the restriction cannot even be motivated by preserving classical mereology.) The removal of this restriction allows for interesting generalizations. In this sense, the route to a theory of infinitesimal regions that provides an attractive account of vectors converges with a natural generalization of Einstein algebras.

Just as Einstein algebras can represent smooth manifolds in virtue of \textsc{Manifold-Algebra Duality}, the quotient rings in SIG can represent \textit{generalized manifolds}, which includes infinitesimal ones represented by nilpotent rings. Thus understood, SIG is a preliminary theory of generalized manifolds, which can be developed into a full-blown theory. We can envision its full-blown theory because it has a close connection to \textit{synthetic differential geometry} and its associated \textit{smooth infinitesimal analysis}, as well as some possible variants (such as in Bunge et.al. 2018).  I will turn to this connection in the next section.

It is worth clarifying that SIG can be made compatible with both spacetime algebraicism and substantivalism. As it currently is, SIG is noncommittal in whether spacetime as a substance is fundamental. To make SIG a view of spacetime algebraicism, we can interpret rings as consisting of \textit{physical field configurations}, and interpret \textsc{Ring Representation} as describing how spacetime is \textit{constructed} or derived from fundamental physical fields (see Bain 2003; Norton (2008) also calls spacetime algebraicism ``constructivism''). Such an approach would not undermine the merits of SIG recommended because the questions we are interested in do not depend on the fundamentality of substantive spacetime. For example, without substantive spacetime, fundamental vectorial quantities would be considered intrinsic to infinitesimal parts of fundamental physical fields.

\section{Interpreting Smooth Infinitesimal Analysis}

\textit{Smooth infinitesimal analysis} (SIA), which is part of the more comprehensive theory called  \textit{synthetic differential geometry} (SDG), is an influential theory of infinitesimals that purports to regiment physicists' heuristic reasoning with infinitesimals. However, there is a long-standing problem for interpreting it realistically as a theory of space: it is formulated in intuitionistic logic and is classically inconsistent (see Bell 2008). Also, Hellman (2006) argued that the theory cannot be reconstructed as a classically consistent theory of infinitesimals. But I will argue that SIG can be considered as (part of) a realistic interpretation of SIA (or SDG).  In particular, I will explain how the model consisting of quotient rings can be used to interpret two classically inconsistent statements under \textit{KC semantics} as an illustration of the full-blown semantics (called ``sheaf semantics'') and models proposed by Moerdijk and Reyes (1991). Then, I will recommend adopting the semantic view of SIA and taking its models realistically.

SIA features the indefinitely extended smooth line $\mathcal{R}$ that carries usual operations like addition and multiplication (that is, $\mathcal{R}$ is a field). The idea that an infinitesimal segment of a curve is straight and non-degenerate (i.e., not identical to a point) is formally captured by a core principle called the Kock-Lawvere axiom (``$f$'' ranges over smooth functions; $\Delta=\{x\in\mathcal{R}\mid x^2=0\}$):

\begin{quote}
	\textsc{Kock-Lawvere Axiom (KL).} $(\forall f:\Delta\to \mathcal{R}) (\exists! a,b\in \mathcal{R}) (\forall x\in \Delta) f(x)=a+bx.$
\end{quote}
We say that all smooth functions on $\Delta$ are \textit{affine}. KL implies that, for any smooth function on $\mathcal{R}$, when it is restricted to an area as small as $\Delta$, its graph becomes straight. Note that this is very similar to the features of nilpotent regions (Section 2.2). SIA is formulated in intuitionistic type theory (Moerdijk and Reyes 1991; see also Feferman 1985).\footnote{In Feferman's term, it is a language of ``variable types'': there are not only types to which each term belong, there are also variables for types. Thus we can talk about properties of types in such a language.} Here, the variable $f$ is of a function type, while $\mathcal{R}$ and $\Delta$ are number types.\footnote{It is worth remembering that SIA conflicts with standard set theory: as I will explain soon, the set of nilpotent infinitesimals $\Delta$ is not the singleton of zero, but it doesn't have other members. So the function type $\Delta \to \mathcal{R}$ is not reducible to pairs of numbers respectively from $\Delta$ and $\mathcal{R}$. } Otherwise, the language of SIA is similar to that of standard analysis. For example, there are constants for real numbers, and function symbols for operators like $+,\cdot, \log$ and relation symbols like $=$ and $<$.

 KL implies that not all nilpotent numbers are zero (i.e., $\Delta\neq \{0\}$). For suppose $\Delta=\{0\}$; then any function $f$ over $\Delta$ would be a constant function, and so there would not be a unique $b$ such that $f(x)=a+bx$, which contradicts KL. However, although $\Delta$ is not identical to $\{0\}$, we can nevertheless prove that no element of $\Delta$ is distinct from zero. Suppose there is a non-zero nilpotent infinitesimal $\epsilon$. Given that $\mathcal{R}$ is a field, which implies that non-zero elements of $\mathcal{R}$ have multiplicative inverses, we have $\epsilon=\epsilon\cdot 1=\epsilon\cdot(\epsilon\cdot \epsilon^{-1})=\epsilon^2\cdot\epsilon^{-1}=0\cdot\epsilon^{-1}=0.$ This contradicts the assumption that $\epsilon$ is not zero. So there are no non-zero nilpotent infinitesimals.\footnote{Note that this proof is valid in intuitionistic logic. Although we cannot prove a statement to be true by showing its negation leads to contradiction in intuitionistic logic, we can nevertheless derive the negation of a statement by showing that the statement leads to a contradiction.
	
	We can also prove that nilpotent infinitesimals are not invertible. Suppose $\epsilon$ is an invertible nilpotent infinitesimal. Then we have $1=1\cdot 1=(\epsilon\cdot \epsilon^{-1})\cdot (\epsilon\cdot \epsilon^{-1})=\epsilon^2\cdot \epsilon^{-2}=0$---contradiction. Therefore, nilpotent infinitesimals are not invertible. The result that nilpotent infinitesimals are not non-zero also straightforwardly follows from this and the fact that every non-zero number is invertible. See Shulman (2006). Although infinitesimals are not invertible, they satisfy the cancellation law: if $a\cdot \epsilon=b\cdot\epsilon$, then $a=b$, where $a,b$ are real numbers and $\epsilon$ is an infinitesimal. This is all we need for calculus. See Giordano (2010) for related discussions.}

We have arrived at the following two classically contradictory statements in SIA:
\begin{quote}
	\textsc{Claim 1}.  $\neg \forall x \in \mathcal{R}(x^2=0\to x=0)$.
	
	\textsc{Claim 2}. $\neg\exists x\in \mathcal{R}(x^2=0\wedge x\neq 0)$.
\end{quote}
These claims are nonetheless consistent in intuitionistic logic:  although $\neg \exists x\in \mathcal{R}(x^2=0\wedge x\neq 0)$ intuitionistically implies $\forall x \in \mathcal{R}(x^2=0\to \neg x\neq 0)$, $\neg x\neq 0$ does not intuitionistically imply $x=0$.

Assuming that a realistic theory needs to be classically consistent, we need to reconcile SIA with classical logic in order to take advantage of it as a realistic theory of space. However, this is not easy. Having examined all the options, Hellman comes to ``no clear resolution" (643) on whether there is a classical reinterpretation of SIA. As a result, he does not consider SIA as ``a theory of actual constitution of ordinary space.'' (645) So it seems that we face an unpleasant choice: we either give up on SIA as a realistic theory (Hellman 2006, Reeder 2015) or give up classical logic (Heller and Kr\'ol 2016). However, I will argue that we have a third option---we can interpret SIA as a classically consistent theory of space. Indeed SIG is (part of) such an interpretation. 

The claim that quotient rings constitute part of a model for SIA is not technically new---such classical models are proposed by Moerdijk and Reyes (1991). In what follows, I will illustrate how we can use the quotient rings to interpret the statements of SIA---in particular, the two classically inconsistent claims---and how that helps meet the challenge of interpreting SIA as a realistic theory of space. 

\paragraph{KC Semantics}
The interpretations of SIA proposed by Moerdijk and Reyes in their models require a rich semantics called \textit{sheaf semantics} (see Mac Lane and Moerdijk 1992). But since I will only show the interpretation of \textsc{Claim 1} and \textsc{Claim 2} from the last section, which involve only one type $\mathcal{R}$, I will appeal to a much simplified semantics, which is Kripke semantics for intuitionistic logic with counterpart relations---call it \textit{KC semantics}.  I will explain the semantics in set-theoretical language (as opposed to the language of category theory, which is typically used for sheaf semantics).  A model under KC semantics is a triple $\langle W, C, \sigma\rangle$. $W$ is a set of objects analogous to ``possible worlds.'' $C$ is a set of maps between members of $W$ that determine ``counterpart relations'' between things in those worlds. $\sigma$ is an interpretation function that, for each $w$ in $W$, assigns objects of $w$ to constants, $n$-tuples to $n$-ary predicates, and single-valued $n$-tuples to ($n$-1)-ary function constants.  The main difference between a KC model and a Kripke model is that, instead of a single accessibility relation between possible worlds, there can be many (often infinitely many) counterpart maps between worlds. Indeed, a Kripke model can be considered as a KC model with at most one counterpart map between any two possible worlds.

\paragraph{The Basic Model}
In what I call \textit{the basic model} for SIA, $W$ is a set of quotient rings of $C^\infty(\mathbb{R})$, and the counterpart maps in $C$ are all the homomorphisms among them. For convenience, we can define the accessibility relation between worlds: for any rings $w,w'$, we define that $w'$ is \textit{accessible from} $w$ (abbr.\@ $wRw'$) just in case there is a homomorphism from $w$ to $w'$. It follows that the accessibility relation is reflexive and transitive just like in Kripke semantics for intuitionistic logic.

For every real number constant $c$ in SIA, the interpretation $\sigma$ assigns to $c$ the constant function \textbf{c} in each member of $W$. The addition and multiplication in SIA corresponds to addition and multiplication on smooth functions in each member of $W$.\footnote{In a full-blown model for SIA, the rings of smooth functions are equipped with every operator that can be used on real numbers, which can be used to interpret every operator in SIA. Such rings are called ``$C^\infty$-rings.'' (Moerdijk and Reyes 1991)} For example, ``$0+1$'' is interpreted as $\textbf{0+} \textbf{1}$ (where \textbf{0} and \textbf{1} are the constant functions of value zero and one, and \textbf{+} is the addition of smooth functions). If $P$ is an atomic sentence, for each $w$, $\sigma_w(P)$ is either 1 (``true'') or 0 (``false'') determined in the usual way. For instance, $``0+1=1"$ is true for every $w$ since in every ring $\textbf{0+} \textbf{1}$ is equal to $\textbf{1}$.

For logically compound formulas in SIA (restricted to first-order sentences that have only variables of type $\mathcal{R}$), the clauses for conjunction, disjunction, and the existential quantifier are like in standard semantics for first-order logic.\footnote{Apart from having only one type, the difference between KC semantics and more general sheaf semantics is this: when unpacking the existential quantifier or the connective ``or" at a possible world $w$ in sheaf semantics, it involves evaluating the relevant sentential components not only at $w$ but also all the possible worlds in an \textit{open cover} of $w$, an additional structure of a model. Thus, KC semantics can be considered as a special sheaf semantics with the constraint that every world has itself as the sole member of its open cover. (See Moerdijk and Mac Lane 1992 for the full detail.) A full interpretation of SIA proposed by Moerdijk and Reyes (1991) involves full-blown sheaf semantics. But all the statements of SIA examined in this paper can be interpreted through presheaf semantics alone. Therefore, I will henceforth not worry about the difference between the two semantics.} The following clauses are more special:

\begin{quote}
	\textsc{Negation.} $w\models \neg \phi[s]$ iff for any $w'$ such that $wRw'$ and any $f$ from $w$ to $w'$, $w'\not\models \phi[f\circ s]$.
	
	\textsc{Conditional.} $w\models \phi\to \psi[s]$ iff for all $w'$ such that $wRw'$,  for any  $f$ from $w$ to $w'$, if $w'\models \phi[f\circ s]$, then $w'\models \psi[f\circ s]$.
	
	\textsc{Universal Quantifier.} $w\models \forall x\phi[s]$ iff for all $w'$ such that $wRw'$,  and for all $d\in w'$, $w'\models \phi[ s (x\mapsto d)]$.\footnote{``$s(x\mapsto d)$'' means an assignment like $s$ except assigning $d$ to variable $x$.}
	
\end{quote}
A sentence $S$ is a theorem in SIA if for every $w\in W$ and every assignment $s$, we have $w\models S[s]$. Notice that the above connectives behave as if they were in the scope of ``necessarily" in modal logic with counterpart relations. For example, the clause for ``$\neg p$" is like the classical clause for ``necessarily, not $p$." Thus, $p\vee\neg p$ is not a theorem.

\paragraph{Interpreting SIA}
First, consider \textsc{Claim 1}: $\neg \forall x \in \mathcal{R}(x^2=0\to x=0).$	Its interpretation is as follows: for any rings $w_1, w_2$ with $w_1Rw_2$, it is not the case that, for any ring $w_3$ with $w_2Rw_3$, for all $d\in w_3$, for any ring $w_4$ with $w_3Rw_4$ and any $f$ from $w_3$ to $w_4$ if $f(d)^\textbf{2}=\textbf{0}$, then $f(d)=\textbf{0}$. (The $w_1$ quantifier comes from evaluating the sentence at every ring; the $w_2$ quantifier comes from the negation; the $w_3$ quantifier comes from the restricted quantifier; and the $w_4$ quantifier comes from the conditional.) This boils down to the condition that every ring has access to some ring that has non-zero nilpotent elements. This is true in the basic model because the ring of affine functions $\mathfrak{L}$ is accessible from every quotient ring of $C^\infty(\mathbb{R})$.\footnote{For any quotient ring of $C^\infty(\mathbb{R})$, there is a homomorphism that maps each ring element $A$ to an element of $\mathfrak{L}$ by evaluating the value and derivative of a representative member (arbitrarily chosen) of $A$ at a point that all members of each element of the quotient ring agree on (for any nonzero quotient ring of $C^\infty(\mathbb{R})$, there is at least one point that all members of each element agree on).}

Consider \textsc{Claim 2}: $\neg\exists x\in \mathcal{R}(x^2=0\wedge x\neq 0).$
The interpretation of this claim in the model is that, for any rings $w_1, w_2$ with $w_1Rw_2$, it is not the case that, there is a $d\in w_2$ such that $d^\textbf{2}=\textbf{0}$ and for any ring $w_3$ with $w_2Rw_3$ and any map $f$ from $w_2$ to $w_3$, $f(d)\neq \textbf{0}$. (The $w_2$ quantifier comes the first negation and the $w_3$ quantifier comes from the second negation.) This comes down to the condition that every nilpotent element has zero as a counterpart, which is true in the model. For every quotient ring of $C^\infty(\mathbb{R})$, there is a homomorphism from it to $\mathbb{R}$, which simply maps every ring element to its value at zero.  The only nilpotent element in $\mathbb{R}$ is 0. Since a homomorphism must preserve squares, any nilpotent element in any ring is mapped to 0 in $\mathbb{R}$ under any homomorphism. So, every nilpotent element in every quotient ring has zero as a counterpart.

\paragraph{}

The basic model is a substructure of a full-blown model for SIA. Like what we see for \textsc{Claim 1} and 2, all statements of SIA can be interpreted through sheaf semantics as statements about quotient rings of $C^\infty(\mathbb{R})$ and their relations. Usually, such a model is considered an abstract tool for proving the consistency of SIA. But we can reverse this order: we can take the object language instrumentally and the models realistically.\footnote{It might seem odd to some to take SIA as a heuristic device for reasoning about the models, because it is not so convenient to use intuitionistic logic. To reply, I shall point out that it is still easier to derive statements in SIA than reasoning about the its models directly, especially its full-blown models that involve sheaf structures (see Moerdijk and Reyes 1991). Also see Footnote 29.} Akin to the semantic view of theories in van Fraassen (1980), we can consider SIA as saying that one of its classical models corresponds to reality. 

But this can't be the whole story. An interpretation should also preserve the important characteristics and virtues of the theory to be interpreted (or ``discharge its scientific duties,'' as Ruetsche 2011 puts it). One apparent obstacle for taking the models seriously is that the models seem to be about smooth functions on real coordinate space: we can consider these smooth functions as representing physical scalar fields living on standard space and have certain algebraic properties. This does not seem to say anything substantially different from the standard view of space, and therefore does not capture what SIA is about. However, this reading is not mandatory. We can interpret the rings according to \textsc{Ring Representation} and other structures accordingly, and the resulting theory is precisely SIG (or rather, its fully developed versions). As we have seen, SIG preserves important claims of SIA such as space has infinitesimal regions on which all smooth functions are affine, which are responsible for its characteristic virtues of regimenting a mode of reasoning with infinitesimals and providing a simple metaphysical analysis of vectors and tangent space. This justifies treating SIG as a realistic interpretation of SIA, and SIA a heuristic device for SIG.\footnote{Despite SIG involving a much simplified version of the full models proposed by Moerdijk and Reyes, we can see that it is still complicated to obtain even the most basic results (see Section 2.1). Thus, although SIA is less straightforward than standard analysis, it's still easier to use than SIG. Also see Footnote 28.}

One may object that although SIA and SIG share many similarities, the translation from SIA to SIG through the semantics does not seem to be even approximately faithful. For example, the claim that not all nilpotent numbers are zero is translated into the algebraic relations between different rings through KC semantics. Even if we understand those rings geometrically through the lens of SIG, the claim would still be about the geometric relations between different regions (including a nilpotent region), which significantly deviates from what the original statement says. To reply, I shall note that it is an open question whether there is a more direct way of translating from SIA to SIG through the exchange between intuitionistic logic and nonclassical mereology.\footnote{The sheaf models for SIA proposed by Moerdijk and Reyes are \textit{topoi}. We can embed the basic model into the sheaf model in a way that preserves all homomorphisms between all rings in the basic model (except that all directions are reversed), known as \textit{Yoneda embedding}. Here, we interpret the rings as representing regions of space, and homomorphism as representing smooth functions between regions (except that all directions are reversed). This is roughly why there is a structural similarity between regions of space in SIG and the subsets of $\mathcal{R}$ that SIA talks about. One main reason why I can appeal to a much simpler model than the sheaf models is that the language I use to describe the theory is not definable in the basic model, while the language of SIA (intuitionistic type theory) is internal to the sheaf model (all elements of the language, including logic symbols, can be defined in the topoi). If we can formulate the internal language of a topos as one with nonclassical mereology rather than nonclassical logic, then we can more directly reformulate SIA into a classically consistent theory that has nonclassical mereology. Note that for such a theory---because it is modeled in a topos, which involves complicated superstructures on rings---would not have a simple and direct relation with rings as in \textsc{Ring Representation}.} But regardless of the answer,  the strategy at least opens up a new interpretative option for those who are enticed by the virtues of SIA as a theory of space but do not want to give up classical logic, and are comfortable with reasoning in SIA. They can turn to SIG as a realistic theory of space while continue to use the SIA as a reasoning tool.

\section{Conclusion}

I have proposed a new theory of space called ``smooth infinitesimal geometry'' (SIG) according to which space has infinitesimal regions that are too small to ``bend'' (that is, on which all smooth functions are affine). This theory can regiment a convenient and elegant mode of scientific reasoning with infinitesimals, and provide a straightforward strategy for tackling philosophical puzzles about vectorial quantities and tangent space. It is also connected to the literature on spacetime algebraicism (according to which physical fields exist without a fundamental spacetime manifold), and can be considered as a natural generalization of Einstein algebras (a standard implementation of spacetime algebraicism). Moreover, SIG can be considered a realistic interpretation of smooth infinitesimal analysis, which shares many features with SIG but is formulated in intuitionistic logic and is classically inconsistent, a feature that has concerned many philosophers.   On the other hand, SIG has highly unorthodox features: its mereology is nonclassical and the principle of supplementation fails. In particular, an infinitesimal region has a point as proper part without a remainder. The overall benefit of such a theory is left to the readers to evaluate. To develop a complete theory of space from here, there are many more questions to consider, such as how we should understand spacetime curvature and Einstein field equations. These are left for other occasions.

\end{document}